\documentclass[amsmath,floatfix,prb]{revtex4}
\usepackage{graphicx}
\begin{document}

\title{ Linear and Second-order Optical Response of the III-V Mono-layer Superlattices } 

\author{S. Sharma}
\email{sangeeta.sharma@uni-graz.at}
\author{J. K. Dewhurst}
\author{C. Ambrosch-Draxl}
\affiliation{Institute for Theoretical Physics, Karl--Franzens--Universit\"at Graz,
Universit\"atsplatz 5, A--8010 Graz, Austria.}

\date{\today}

\begin{abstract}

We report the first fully self-consistent calculations of the nonlinear optical properties of superlattices. The materials 
investigated are mono-layer superlattices with GaP grown on the the top of InP, AlP and GaAs (110) substrates. 
We use the full-potential linearized augmented plane wave method within the generalized gradient approximation to obtain 
the frequency dependent dielectric tensor and the second-harmonic-generation susceptibility. The effect of lattice 
relaxations on the linear optical properties are studied. Our calculations show that the major anisotropy in the 
optical properties is the result of strain in GaP. This anisotropy is maximum for the superlattice with maximum lattice mismatch 
between the constituent materials. In order to
differentiate the superlattice features from the bulk-like transitions an improvement over the existing effective medium model is 
proposed. The superlattice features are found to be more pronounced for the second-order than the linear optical response
indicating the need for full supercell calculations in determining the correct second-order response. 

\end{abstract}

\maketitle

\section{INTRODUCTION}


Semi-conducting strained superlattices (SLs) are potential materials for applications in optical communications involving switching, 
amplification and signal processing. In particular III-V semiconductor hetero-structures and SLs have attracted a great deal of interest 
mainly due to the possibility of tailoring band gaps and band structures \cite{osbourn83,MBEH85,pearsall87,IQWS88,BSESMJ89} by variation 
of simple parameters like superlattice period, growth direction and substrate material. With the development of new techniques like the 
strain induced  lateral ordering process \cite{mascarenhas93} and existing methods like molecular beam epitaxy \cite{HBDPDA87} and low-pressure 
chemical-vapor deposition\cite{angus88} it is possible to grow and tailor these SLs. Thus a great deal of experimental work has 
been devoted to these materials. The unusual optical behaviour of AlP/GaP SLs has been extensively studied using 
photoluminescence, magneto-photoluminescence, \cite{uchida98,issiki96,nabetani95} optical absorption,\cite{wang95} 
X-ray diffraction\cite{xue94} and refractive index measurements.\cite{morii94}  
InP/GaP, being one of the material combinations which spontaneously constructs the SL under specific conditions, has been subject to numerous experimental 
works \cite{wang91,gomyo88,cheong99} including cathodoluminescence experiments \cite{rich97,tang97,tang96} and studies on the influence of  
pressure, SL period and barrier thickness \cite{fudeta99,kim98,seong98} on the optical transitions. Likewise GaAs/GaP \cite{zhou92,recio90} 
and GaAs/AlAs\cite{chavez92,shih94,nguyen93,blom93,pusep93,vercik02,sun00} have also been extensively investigated in the past. 
 
Much of the  theoretical work done to explain these interesting physical properties of SLs has been largely concerned with 
the understanding of the electronic band structure. For example, the effect of strain on the band gap, the band offset problem 
and the possibilities of engineering it as well as the interface energy and band structure 
have been studied. \cite{franceschetti99,walle86,chris87,munoz90,agrawal99,dandrea91,tanida94,park93,kurimoto89,arriga91} 
The linear optical properties of SLs have also been determined theoretically. Franceschetti {\it et al.} have calculated the pressure coefficients 
of optical transitions in InP/GaP SLs. \cite{franceschetti94} The pseudopotential method within the local density approximation (LDA) has been 
used by Kobayashi {\it et al.}\cite{kobayashi96} to calculate the optical transition strengths of the (AlP)$_n$/(GaP)$_m$ type multi-quantum-well SLs
and for GaAs/AlAs by Yee {\it et al.}\cite{yee93} Confinement effects on optical transitions for various superlattice periods in GaAs/AlAs 
has been discussed by Schmid {\it et al.}\cite{schmid92} The $sp^3s*$ tight-binding method has been used for the study of optical properties and 
the indirect-to-direct band gap transition of AlP/GaP \cite{kumagai87,kumagai88} and GaAs/AlAs SLs.\cite{tit98} Botti {\it et al.}\cite{botti01} 
have employed the linear combination of bulk bands method to determine the optical properties of GaAs/AlAs SLs and electroabsorption properties of 
these SLs have been studied by Kawashima {\it et al.} \cite{kawashima98} Shibata {\it et al.} \cite{shibata94} used the pseudopotential method to 
determine the oscillator strength and band structures of the AlP/GaP SLs. 

Due to the simultaneous breaking of the time reversal and inversion symmetries at surfaces and interfaces, the nonlinear optical properties are 
more sensitive than the linear optical properties and so the second harmonic generation (SHG) by some of these SLs has also been calculated and 
determined experimentally. \cite{tang96} The major theoretical work in this direction was done by Ghahramani {\it et al}.
\cite{ghahramani92,ghahramani91-si,ghahramani90,ghahramani91} They used the effective-medium-model (EMM) to determine the linear and nonlinear
optical properties of  (GaAs)$_n$/(GaP)$_n$, (Si)$_n$/(Ge)$_n$ and (GaAs)$_n$/(AlAs)$_m$ SLs. They also employed the non-self-consistent 
linear-combination-of-Gaussian-orbitals (LCGO) method within the LDA to calculate the band structures and optical properties of these compounds. 
In these works Ghahramani {\it et al.}  conclude that away from the absorption edge, most of the features in the linear as well as nonlinear 
optical properties are due to the bulk transitions which are well modeled by the EMM. 

There have been very few studies of the linear optical properties of SLs and even fewer calculations of the nonlinear properties. Also the existing 
theoretical work is neither self-consistent nor takes into account the effect of lattice relaxations on the SL optical properties. Hence, there is a need for 
a fully self-consistent calculation of the nonlinear and linear optical properties of superlattices in conjunction with a study of the
lattice relaxation effects. The aim of the present work is to calculate the linear and the nonlinear optical properties of III-V mono-layer SLs 
using the state-of-the-art full-potential linearized augmented plane wave method (FPLAPW). The materials investigated  are (InP)$_1$/(GaP)$_1$, 
(AlP)$_1$/(GaP)$_1$ and (GaAs)$_1$/(GaP)$_1$. The EMM is also tested for all these materials. For InP/GaP and  AlP/GaP, this model 
fails to reproduce even the features away from the absorption edge. So the EMM is reanalyzed and a modification is proposed. 

The paper is arranged in the following manner. In Section II we present the details of the calculations. Sections IIIA and IIIB deal with the 
linear and second-order optical response of mono-layer SLs, respectively. Section IV provides a summary of our work. Appendix A lists all 
the formulae used for the calculation of the SHG susceptibility. Details of the modifications to the EMM are presented in Appendix B.

\section{METHODOLOGY}

Total energy calculations are performed using the FPLAPW method implemented in the {\sf WIEN2k} code.\cite{WIEN} 
Thereby the scalar relativistic Kohn-Sham equations are solved in a self-consistent scheme. For the exchange-correlation potential we use the 
generalized gradient approximation (GGA) derived by Perdew and Wang. \cite{perdew92} 

The  detailed formalism for the determination of the linear dielectric tensor $\epsilon(\omega)=\epsilon_1(\omega)+i\epsilon_2(\omega)$ 
within the FPLAPW formalism has been presented before.\cite{cad} The susceptibility for the second harmonic 
generation $\chi^{(2)}(2\omega,\omega,\omega)$ has been calculated using the an extension to this program.\cite{NLO} The formulae for calculating 
$\chi^{(2)}(2\omega,\omega,\omega)$ have also been presented before.\cite{sipe96,sipe93,sipe00,segey98}  They are rewritten to improve 
the computational efficiency, and we compare the computing time for both sets of formulae in Appendix A. The SHG susceptibility obtained satisfies
all the theoretical sum rules presented by Scandolo and Bassani.\cite{scandolo95}

All the calculations are converged in terms of basis functions as well as in the size of the $k$-point mesh 
representing the Brillouin zone. The linear optical properties are calculated on a mesh of 500 $k$-points in the irreducible Brillouin zone 
(IBZ) and for the second-order susceptibility a mesh of 1500 $k$-points in the IBZ is used. 

\begin{table}[h]
\caption{Superlattice lattice parameters in \AA ~ units. $a_\parallel$=$a_{x}$=$a_{y}$ and $a_\perp$=$a_{z}$ which is the 
direction of crystal growth.}
\centerline{
\begin{tabular}{|c| c | c | } \hline
Superlattice        & $a_\parallel$  &  $a_\perp$ \\ \hline
(InP)$_1$/(GaP)$_1$ &   4.150        &  5.660     \\ \hline 
(AlP)$_1$/(GaP)$_1$ &   3.933        &  5.457     \\ \hline 
(GaAs)$_1$/(GaP)$_1$&   3.998        &  5.553     \\ \hline 
\end{tabular}
}
\end{table}

In all the superlattices GaP is grown on top of other group III phosphates (InP and AlP) or GaAs as the substrate material. 
The superstructures are constructed in the (110) direction, which is equivalent to (100) direction in these materials. All these structures are treated in 
tetragonal unit cells. The details of 
this unit cell for such SLs has been discussed before (Fig. 1(a) in Ref. \onlinecite{ciraci88}). The lattice constants of the SL are 
estimated using the macroscopic elasticity theory\cite{arriga91,people85,walle86} and are presented in Table I. The experimental lattice
constants of the constituent materials (InP, GaP, AlP and GaAs) are taken from Ref. \onlinecite{huang93}.

\section {RESULTS AND DISCUSSION}

\subsection{LINEAR OPTICAL RESPONSE}

First, we shall discuss the importance of lattice relaxations and their effect on the linear optical properties. Total energy and 
force calculations performed for the unrelaxed atomic positions of the mono-layer InP/GaP and AlP/GaP SLs show considerable forces on the atoms, 
so the relaxation of the structure is needed.  On complete relaxation there is a gain in energy of 0.195 and 1.124  eV per formula-unit
for (InP)$_1$/(GaP)$_1$ and (AlP)$_1$/(GaP)$_1$, respectively. For the case of (GaAs)$_1$/(GaP)$_1$ the atomic relaxations were not needed.


\begin{figure}[ht]
\centerline{
\includegraphics[scale=0.5,angle=0]{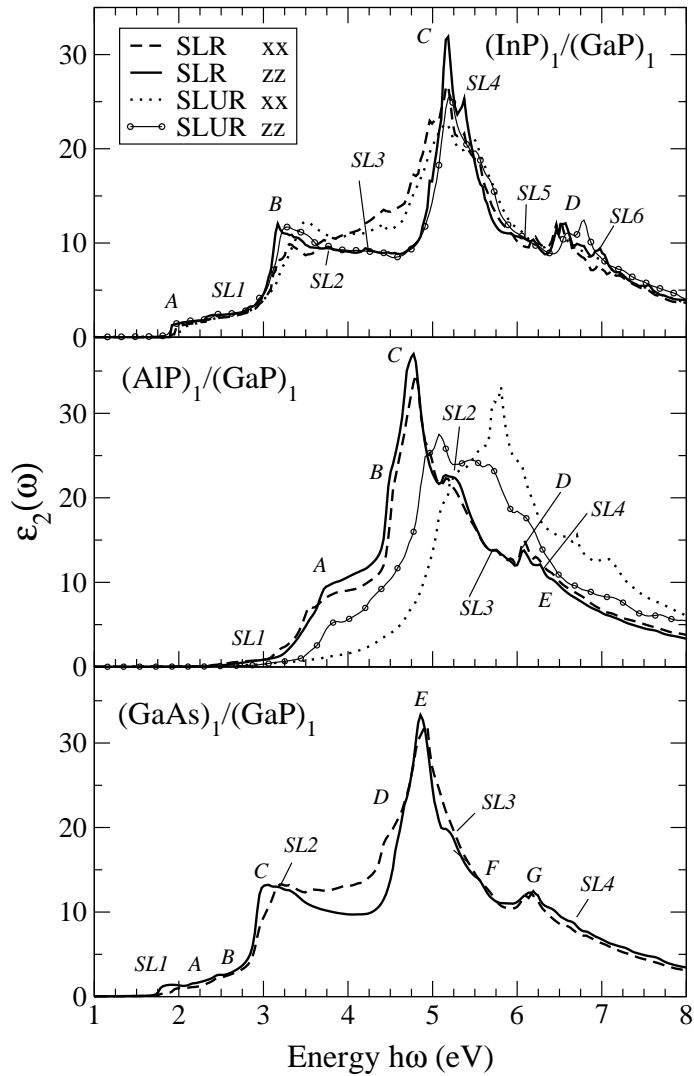}
}
\caption {Components of the imaginary part of the dielectric tensor for the relaxed (SLR) and un-relaxed (SLUR) 
superlattice for (a) (InP)$_1$/(GaP)$_1$, (b) (AlP)$_1$/(GaP)$_1$ and (c) (GaAs)$_1$/(GaP)$_1$. The important features
in $\epsilon_2^{zz}(\omega)$ for the relaxed SLs are labeled with the uppercase letters.}
\end{figure}

The imaginary part of the dielectric function, $\epsilon_2(\omega)$,  for all the SLs under investigation are presented in Figure 1.  
All the optical results presented in this paper are scissors corrected \cite{sipe96} using the scissors operator calculated by taking the
difference between the theoretical and experimental band gap of the SL.
Among the SLs under investigation for the relaxed structures, the anisotropy in $\epsilon_2(\omega)$ is maximum for (InP)$_1$/(GaP)$_1$ 
(Fig. 1(a)), where the lattice mismatch between InP and GaP is 7.38 \% , while for (AlP)$_1$/(GaP)$_1$  (Fig. 1(b)) the anisotropy is 
minimum and so is the lattice mismatch (0.2\%). (GaAs)$_1$/(GaP)$_1$ (Fig. 1(c)) lies in between the two with a lattice mismatch between 
GaAs and GaP of 3.66 \%. These trends indicate that most of the anisotropy is due to the strain in the growth material rather than to 
the lower symmetry of the superlattice. These results are in accordance with similar observations previously made for GaAs/AlAs and Si/Ge 
SLs.\cite{ghahramani91,ghahramani90} On the other hand the anisotropy in $\epsilon_2(\omega)$ of the unrelaxed SLs shows a 
maximum for the SL formed from the materials with minimum lattice mismatch (AlP/GaP). Also $\epsilon_2(\omega)$ of the unrelaxed SLs is lower 
in terms of peak heights compared to the relaxed SLs. Hence, the lattice relaxations are very important in determining the correct anisotropy 
and peak heights and thus all the further calculations are performed for the relaxed SLs. Our results for $\epsilon_2(\omega)$ of (GaAs)$_1$/(GaP)$_1$ 
are qualitatively in reasonable agreement with previous calculations by Ghahramani {\it et al.} \cite{ghahramani92} However, the magnitude of the 
response in the present results is nearly two times that obtained previously. This is expected since in their work Ghahramani {\it et al.} pointed
out that the LCGO method is not able to reproduce the correct magnitude of the optical response and their results could be 
underestimated by a factor of 2.  The other differences in the two theoretical results are as follows: 
(1) The present calculations give one sharp peak around 4.75 eV (labeled {\it E}), while previous results show a broad multiple peak structure 
in $\epsilon_2^{zz}(\omega)$ and $\epsilon_2^{xx}(\omega)$ around the same energy.
(2) In the energy region 3.2 - 4.6 eV  $\epsilon_2^{xx}(\omega)$ is greater in magnitude than $\epsilon_2^{zz}(\omega)$ in the present work, 
where as the previous results show the opposite.
These differences can be due to the difference in the method of calculation. Ghahramani {\it et al.} have used a non-self-consistent LCGO method
within the LDA,  while,  the present calculations are fully self-consistent using the FPLAPW method within the GGA.

\begin{figure}[ht]
\centerline{
\includegraphics[scale=0.5,angle=0]{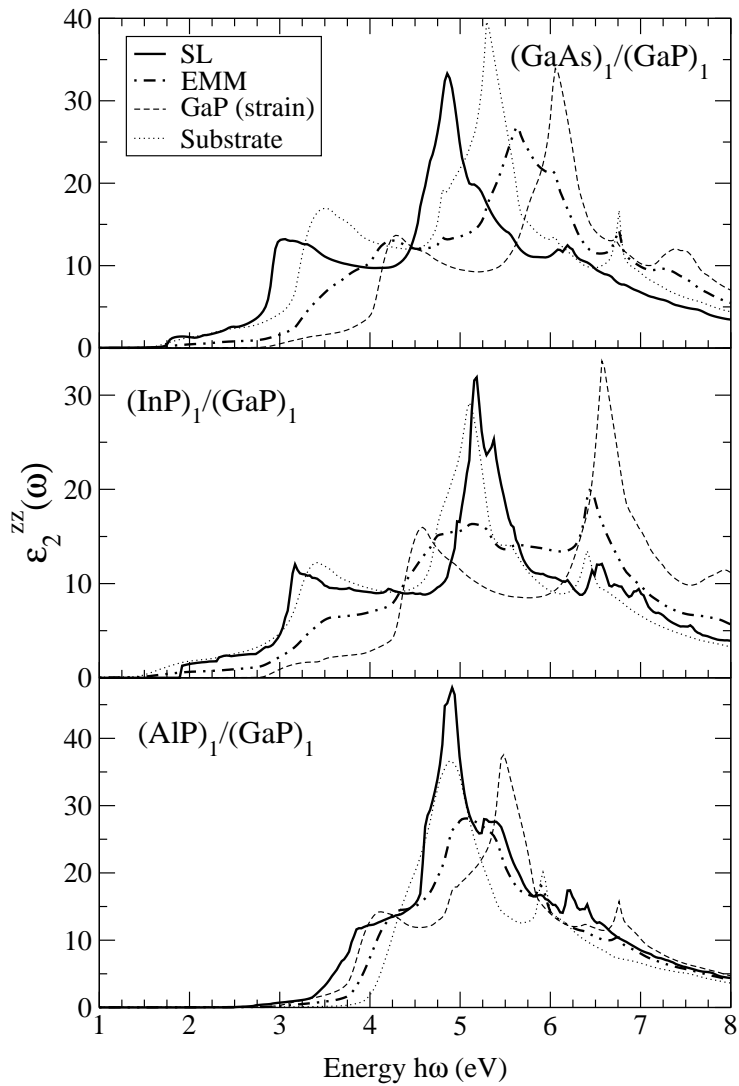}
}
\caption{Frequency dependent $\epsilon_2^{zz}(\omega)$ obtained by the SL calculations compared to the EMM results for (a) (GaAs)$_1$/(GaP)$_1$, 
(b) (InP)$_1$/(GaP)$_1$ and (c) (AlP)$_1$/(GaP)$_1$. 
In addition the data for the bulk GaAs (a), InP(b), AlP(c) and the strained bulk GaP ((a), (b) and (c)) are also presented.}
\end{figure}

A macroscopic model for the dielectric function of SLs has been suggested by Ghahramani {\it et al.} \cite{ghahramani90} In this model the SL is 
considered to be constructed of slabs of an unstrained bulk substrate material (InP, GaAs or AlP in the present case) and a strained bulk material (GaP 
in the present case) grown on top of the substrate. This model is called the effective medium model (EMM). We have used this model to estimate the
macroscopic averaged response and the results are presented in the Figure 2. Like Ghahramani {\it et al.} \cite{ghahramani92} 
we find that the EMM generates reasonably good intensities but poor peak positions in (GaAs)$_1$/(GaP)$_1$ (Fig. 2(a)). In the optical spectrum of 
(InP)$_1$/(GaP)$_1$ (Fig. 2(b)) and (AlP)$_1$/(GaP)$_1$ (Fig. 2(c)) the peak positions are  better but the intensities are less well described. 
Another point that should be noted is that for (AlP)$_1$/(GaP)$_1$, where the lattice relaxations are important, the EMM 
calculated response is closer to that of the unrelaxed SL than to the relaxed SL response, but the same is not true for (InP)$_1$/(GaP)$_1$  
such that no consistent picture emerges.

In order to find the reason for this behaviour, ${\epsilon_2}^{zz}(\omega)$ for bulk InP, GaAs and AlP and strained GaP are also presented in Figure 2. 
We find that if the major peak positions in the dielectric function of the constituent materials are well separated in energy the EMM gives two 
separate peaks. As can be seen for the (InP)$_1$/(GaP)$_1$ the major peak in the optical spectra of InP  and strained GaP are around 5.0 and 6.65 eV 
respectively. The EMM also gives two peaks of almost equal strength around 5.0 and 6.65 eV.  The separation between the major optical peaks of GaAs and 
strained GaP is 0.76 eV and between AlP and strained GaP it is 0.80 eV. The EMM in these cases results in broadened and thus weakened peaks. It is 
surprising that the SL response is seemingly dominated by the substrate material with the growth material having little 
((GaAs)$_1$/(GaP)$_1$) to nearly no contribution ((InP)$_1$/(GaP)$_1$) (Fig. 2).

\begin{figure}[ht]
\centerline{
\includegraphics[scale=0.5,angle=0]{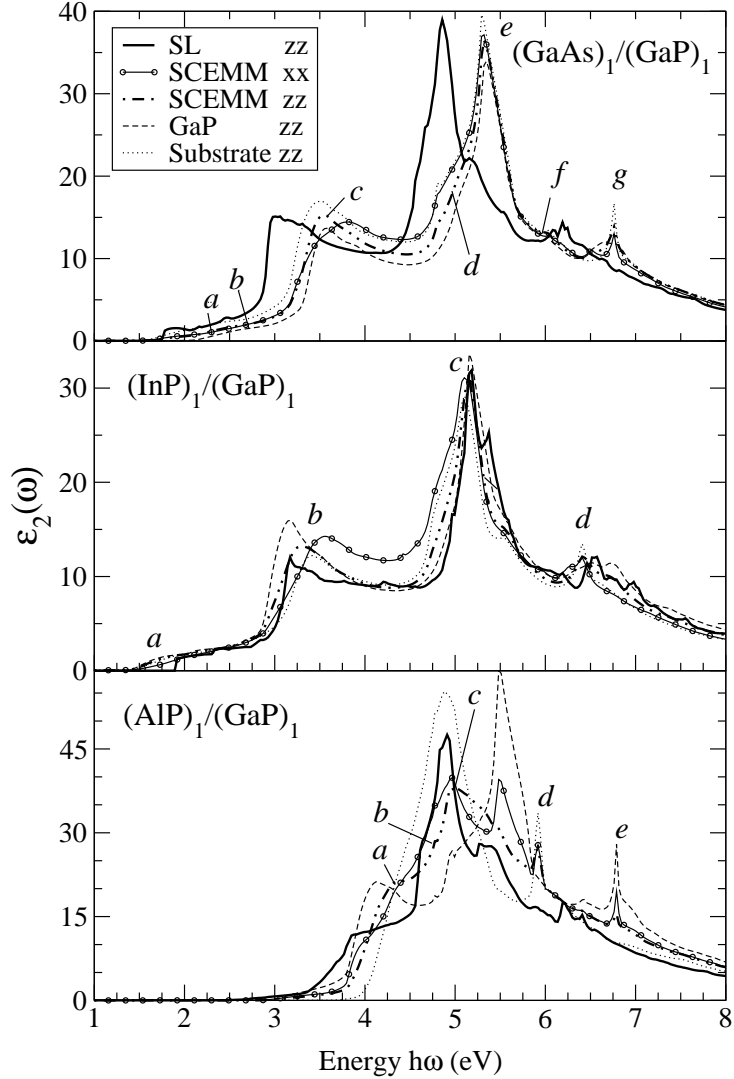}
}
\caption{$\epsilon_2^{zz}(\omega)$ calculated using the full SL and the SCEMM and $\epsilon_2^{xx}(\omega)$ calculated using the SCEMM
for (a) (GaAs)$_1$/(GaP)$_1$, (b) (InP)$_1$/(GaP)$_1$ and (c) (AlP)$_1$/(GaP)$_1$. 
In addition the data for the bulk GaAs (a), InP(b), AlP(c) and the strained bulk GaP ((a), (b) and (c)) are also presented. 
The important features in $\epsilon_2^{zz}(\omega)$ for the relaxed SLs obtained by the SCEMM are labeled with the lowercase letters. }
\end{figure}
 
The purpose of the model is to find the bulk-like features in order to facilitate the detection of true SL features  by comparing the model 
results with the full SL calculations. But EMM takes into account the effect of junction formation only partially by using the 
strained lattice parameters for one of the constituent materials. This is useful for reproducing the anisotropy. The effect of the junction 
formation on the band gap, however, is ignored which is taken to be the experimental gap of the unstrained bulk. As the number of SL layers increases this 
effect diminishes but in the case of mono-layer SLs it is pronounced. So the failure of the model in certain cases is not a surprise.  
Nevertheless, in order to provide a simple model for predicting the bulk-like features in the SL optical properties on the basis of its constituent 
materials, the shortcomings of the EMM can be fixed in the following way: The calculations are performed first for the unstrained growth material 
(GaP in the present case) and the scissors operator is determined by comparison with the experimental data. Now assuming that GGA (or LDA) 
underestimates the gap consistently in the unstrained 
and strained bulk material this same scissors operator is used to correct the calculated linear (and nonlinear) optical response of the strained growth 
material (GaP in the present case). The average of this response with the response of the bulk substrate material (AlP, InP or GaAs in the present case) 
is then taken to determine the effective optical response of the SL. We call this new model the strain-corrected effective medium model (SCEMM). 
The details of the SCEMM are presented in the Appendix B. The results of the SCEMM along with the dielectric function of the substrate and 
strained growth material, corrected by the scissors operator calculated as stated above, are presented in the Figure 3. As can be seen the model 
reproduces most of the features in the SL spectra of all the compounds under investigation. The remaining discrepancy in peak positions is due to the
fact that the experimental SL gap is lower than the average gap of the constituent materials. More importantly the SCEMM shows that the SL response is 
not just dominated by the substrate material, as indicated by the EMM, but has features from both constituent materials. This fact is best seen 
in the case of (AlP)$_1$/(GaP)$_1$ where the SCEMM peak labeled $a$ in Fig. 3(c) is dominated by the strained bulk GaP transitions and $c$ is bulk InP like. 
The anisotropy in $\epsilon_2(\omega)$ determined by the SCEMM confirms 
our earlier observation that most of the anisotropy is due to strain in the growth material rather than to the lower symmetry of the superlattice. 
The features not reproduced by the model are referred to as the SL features in the present work.  
The following features can be identified SL effects: 
$SL1(\sim 2.5eV), SL2(\sim 3.75eV), SL3(\sim 4.25eV), SL4(\sim 5.6eV), SL5(\sim 6eV)$ and $SL6(\sim 7eV)$ in (InP)$_1$/(GaP)$_1$, 
$SL1(\sim 2.85eV), SL2(\sim 5.25eV), SL3(\sim 5.8eV)$ and $SL4(\sim 6.5eV)$ in (AlP)$_1$/(GaP)$_1$ and 
$SL1(\sim 1.88eV), SL2(\sim 3.25eV), SL3(\sim 5.25eV)$ and $SL4(\sim 6.57eV)$ in (GaAs)$_1$/(GaP)$_1$ SLs in Fig. 1

\begin{table}[h]
\caption{ Static dielectric constants $\epsilon^{xx}_1(0)$ and $\epsilon^{zz}_1(0)$ and the components of the SHG susceptibility in the 
static limit $\chi_1^{zyx}(0)$ and $\chi_1^{xyz}(0)$ in $\times 10^{-8}$ esu for the relaxed SLs compared to the SCEMM and the EMM results.}
\centerline{
\begin{tabular}{|c |c c c | c c c |c c c| c c c|} \hline
Superlattice        &      & $\epsilon^{xx}_1(0)$ &       &       & $\epsilon^{zz}_1(0)$   &        &    
                                                                  & $\chi_1^{zyx}(0)$      &        &       &  $\chi_1^{xyz}(0)$     & \\
                    & SL   & SCEMM & EMM   & SL   & SCEMM & EMM           
                                                                   &  SL   & SCEMM & EMM  & SL    & SCEMM  & EMM   \\ \hline
InP/GaP             & 9.27 & 9.60  & 8.51  & 9.00 & 9.30  & 8.17       
                                                                   & 13.89 & 15.57 & 9.47 & 10.42 &  15.55 & 9.29  \\ \hline
AlP/GaP             & 8.65 & 8.25  & 8.27  & 8.86 & 8.28  & 8.30
                                                                   & 6.34  & 6.08  & 6.14 & 7.51  &  6.10  & 6.16  \\ \hline
GaAs/GaP            & 10.48 & 9.76 & 9.23  & 10.19 & 9.54  &  8.93
                                                                   & 16.65 &  11.4 & 9.14 & 15.56 & 11.40  & 8.95  \\ \hline 
\end{tabular}
}
\end{table}

The real part of the dielectric function in the static limit is a directly measurable quantity. Our calculated results of
$\epsilon_1^{xx}(0)$ and  $\epsilon_1^{zz}(0)$ for the relaxed SLs are presented in Table II. Two things should be noted. First the anisotropy 
in $\epsilon_1(0)$ follows the trend of the lattice mismatch with a maximum of 2.96\% for (InP)$_1$/(GaP)$_1$ and a minimum of 2.60\% for 
(AlP)$_1$/(GaP)$_1$ while (GaAs)$_1$/(GaP)$_1$ lies in between the two with an anisotropy in $\epsilon_1(0)$ of 2.80\%. Second, in case of 
(AlP)$_1$/(GaP)$_1$ and (GaAs)$_1$/(GaP)$_1$ the lattice constant for the substrate material (GaAs and AlP) is larger than that of the growth material 
(GaP) and so $\epsilon_1^{xx}(0)$ is greater than  $\epsilon_1^{zz}(0)$, while it is just the opposite for AlP/GaP SL. The results of 
$\epsilon_1(0)$ using the SCEMM are closer to the full SL calculations (worst by a maximum of 6.8\% ) than the results obtained using the EMM 
(worst by a maximum of 12.3\%). A more important test for any model, however, comes in the determination of the nonlinear optical properties as they 
are much more sensitive to the small changes in the bandstructure than the linear optical spectra.

\subsection{SECOND-ORDER OPTICAL RESPONSE}

\begin{figure}[ht]
\centerline{
\includegraphics[scale=0.5,angle=0]{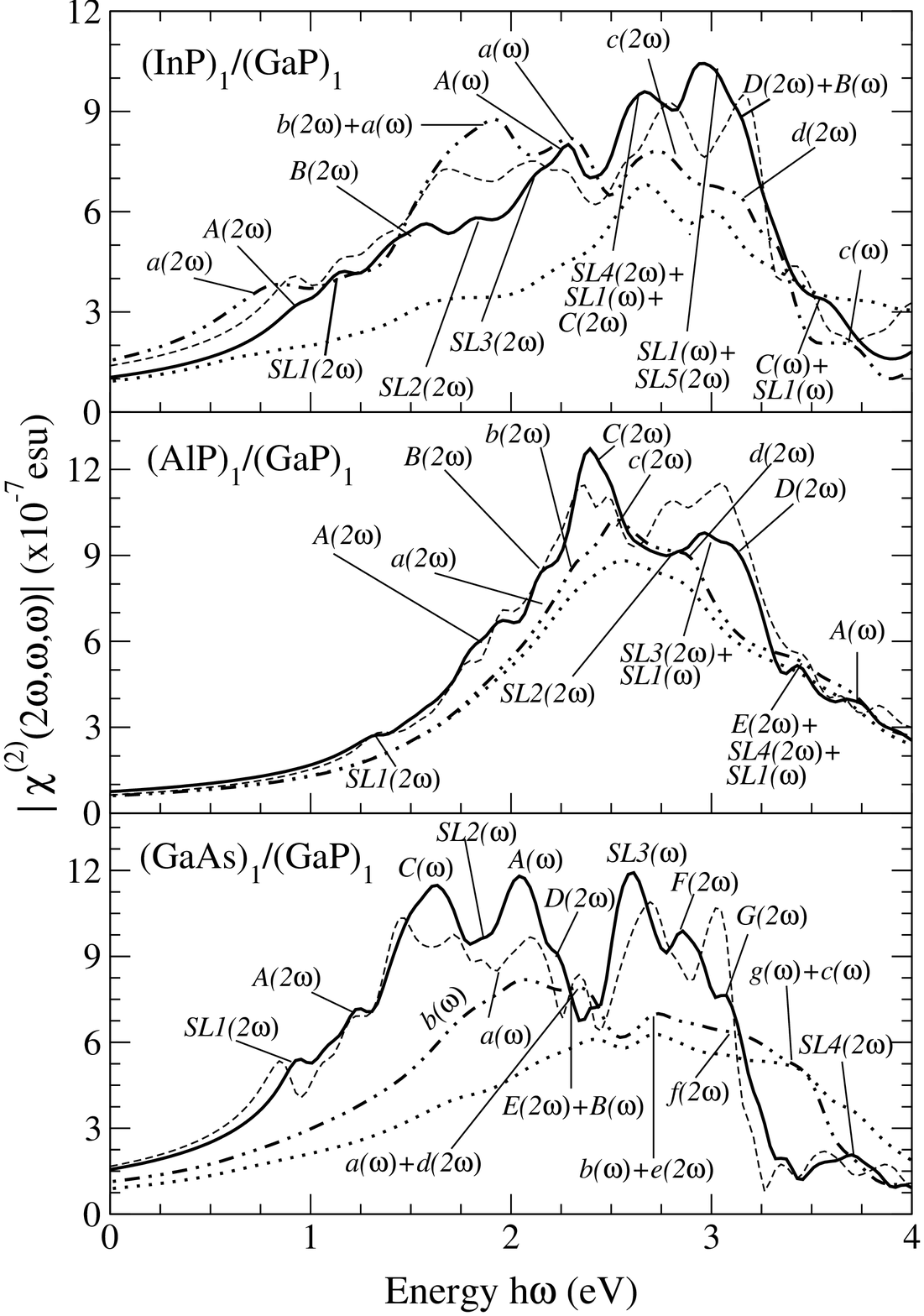}
}
\caption{ 4(a) The two inequivalent components of susceptibility for the SHG $|\chi^{(2)}_{xyz}(2\omega,\omega,\omega)|$ (thick solid line) and 
$|\chi^{(2)}_{zyx}(2\omega,\omega,\omega)|$ (dashed line) obtained from the full SL calculations, along with the SCEMM (thick dash-dot line) and 
the EMM (dotted line) results of $|\chi^{(2)}_{xyz}(2\omega,\omega,\omega)|$ for (InP)$_1$/(GaP)$_1$. 
4(b) The same for (AlP)$_1$/(GaP)$_1$. 
4(c) The same for (GaAs)$_1$/(GaP)$_1$. }
\end{figure}

The magnitude of the SHG susceptibility ${\chi^{(2)}}(2\omega,\omega,\omega)$ for the mono-layer SLs along with the results of the SCEMM and the
EMM are presented in Figure 4. The peaks in ${\chi^{(2)}}(2\omega,\omega,\omega)$ can be identified to be coming from 2$\omega$ and/or $\omega$ 
resonances of the peaks in the linear dielectric function. Therefore this identification of the peaks in the SHG susceptibility is done from the respective 
linear optical spectra. The identified peaks are marked in Figure 4 and the nomenclature adopted is $M(x\omega)+N(y\omega)$, which 
indicates that the peak comes from an  $x\omega$ resonance of the peak $M$ with the $y\omega$ resonance of peak {\it N} in the linear 
optical spectra. For example, for (InP)$_1$/(GaP)$_1$ (Fig. 4(a)) the hump just below 1 eV, labeled $A(2\omega)$ in the 
$\chi^{(2)}_{xyz}(2\omega,\omega,\omega)$ component for the SL comes from the 2$\omega$ resonance of the peak labeled 
{ \it A }in the linear optical spectra (Fig. 1(a)). The peak at 1.15 eV is a SL peak, labeled $SL1(2\omega)$, coming from the 2$\omega$ resonance of the 
peak labeled $SL1$ in the $\epsilon_2(\omega)$ plot (Fig. 1(a)). Similarly all the other features have been identified and marked in Figure 4(a). 
As expected the pure SL peaks $SL1(2\omega)$, $SL2(2\omega)$, $SL3(2\omega)$ and $SL1(\omega) + SL5(2\omega)$ are absent in the SCEMM results. 
The features coming from resonances of the SL and bulk-like peaks such as $SL4(2\omega)+SL1(\omega)+C(2\omega)$ and 
$C(\omega)+SL1(2\omega)$ are underestimated by the SCEMM as features $c(2\omega)$ and $c(\omega)$, while the peaks coming from the 
2$\omega$ and/or $\omega$ resonances of the bulk peaks are well reproduced by the SCEMM. The peak labeled $a(2\omega)$ is more pronounced 
than the corresponding SL peak $A(2\omega)$ in accordance with $a$ (Fig. 3(b)) being larger than $A$ (Fig. 1(a)) in the linear optical spectra. 
In contrast, the peak labeled $B(2\omega)$ is smaller than the corresponding SCEMM peak $b(2\omega)+a(\omega)$ since the SCEMM peak is 
due to the $\omega$ resonance of peak $a$ with the 2$\omega$ resonance of the peak labeled $b$ in the linear dielectric function (Fig. 3(b)).
A similar feature assignment for the (AlP)$_1$/(GaP)$_1$ and (GaAs)$_1$/(GaP)$_1$ mono-layer SLs is also performed and the details are marked in 
Figs. 4(b) and 4(c). The same trends are observed in all the materials under investigation, with the SCEMM reproducing all the peaks other 
than pure SL features.  This indicates that the SCEMM can be used to identify the structure in the optical spectra coming from the bulk-like 
transitions and hence facilitating the determination of the SL effects like symmetry lowering. The EMM on the other hand is not able to reproduce 
all the non-SL features and is therefore not as good a model for the mono-layer SLs. 

Our results for the magnitude of ${\chi^{(2)}}(2\omega,\omega,\omega)$ of (GaAs)$_1$/(GaP)$_1$ (Fig. 4(c)) differ from previous calculations: 
\cite{ghahramani92} 
(1) ${\chi^{(2)}}_{xyz}(2\omega,\omega,\omega)$ is generally in good agreement with the previous work in terms of peak 
heights as well as positions. The only difference is the appearance of two extra features in the present work, i.e., a hump at $\sim$1.25 eV 
($A(2\omega)$) and a peak at 1.75 eV ($SL1(\omega)$).
(2) In the present work the anisotropy in the nonlinear optical response is found to be more pronounced compared to the work of Ghahramani {\it et al.}
As is well known, the nonlinear optical properties are more sensitive to the small changes in the bandstructure than the linear ones so any anisotropy 
in the linear optical response is expected to be enhanced in the nonlinear spectra. This is in accordance with our findings, while the previous 
work shows the opposite. 
(3) Our results for ${\chi^{(2)}}_{zyx}(2\omega,\omega,\omega)$ differ substantially from the previous results, but since a detailed identification of
the peaks is not presented previously it is difficult to point out the discrepancies. As mentioned earlier, these differences could 
be due to the different methods used.

Finally the SHG susceptibility in the static limit is presented in Table II. The maximum anisotropy in ${\chi^{(2)}}(0)$ is in (InP)$_1$/(GaP)$_1$ (28.5\%) 
and the minimum is found for (GaAs)$_1$/(GaP)$_1$ (12.8\%). (AlP)$_1$/(GaP)$_1$ has an anisotropy of 16.8\%. Neither EMM nor SCEMM 
give results close to the SL results. This is in contradiction with the previous findings,\cite{ghahramani92} where Ghahramani {\it et al.} 
report the EMM to give results within 5\% of the SL data. The failure of the SCEMM and the EMM in the present case is not surprising 
since both the models fail to reproduce some of the main features in the optical spectra leading to a substantial difference in the static limit. 
But still the SCEMM results are an improvement over the EMM. 

\section{Summary}

To summarize, we have performed calculations for the linear and nonlinear optical properties of some III-V mono-layer SLs. We conclude the following: 
(1) The lattice relaxations play an important role in the determination of the correct anisotropy and peak heights of the linear optical spectra. 
(2) The effective medium model (EMM) does not reproduce the correct bulk features mainly because it assumes that the optical band gap of the material 
does not change with strain and is taken as the experimental direct gap of the unstrained material. 
(3) An improvement over this model is the strain-corrected effective medium model (SCEMM), which correctly averages the optical properties of the constituent
materials of the SL and facilitates the identification of the SL features in the optical spectra.
(4) The SCEMM results confirm that most of the anisotropy in $\epsilon(\omega)$, in the SLs under investigation, comes 
from the strain in the growth material rather than the symmetry lowering due to SL formation. 
(5) The anisotropy in $\epsilon_2(\omega)$ and in particular in $\epsilon_2(0)$ follow the trend in the lattice mismatch being maximum for 
(InP)$_1$/(GaP)$_1$ and minimum for (AlP)$_1$/(GaP)$_1$. 
(6) The small SL features in $\epsilon_2(\omega)$ are greatly enhanced in the nonlinear spectra and the same is true for the anisotropy. 
(7) The SCEMM is able to correctly reproduce all the bulk-like features in the SHG susceptibility, whereas pure SL features are absent and features coming 
from the resonance of the SL and bulk-like transitions are underestimated. This leads to a failure of the SCEMM (as well as EMM) in the determination 
of ${\chi^{(2)}}(0)$. But the SCEMM is still an improvement over the EMM.  
(8) The SCEMM is good for determining the linear optical properties of the SLs, but for the correct determination of the nonlinear optical
properties SL calculations are essential. The effects of strain due to junction formation and the lattice relaxations are expected to decrease 
with increase in the layer thickness. It would be interesting to compare two models in such a case and to find the SL period for which these 
effects become insignificant. (9) Although no experimental measurements of the SHG susceptibilty for the compounds under investigation  
exist, a detailed comparison of future experimental data with theoretical results would help in the identification of various features in the optical 
spectra, in particular, to highlight the effect of the interface formation in the SLs. This would lead to a better understanding of the physical properties 
which is one of the most essential ingredient for tailorable materials of technological importance.

\section{Appendix A: FORMALISM FOR THE SECOND-ORDER RESPONSE}

The formulae for the total susceptibility for the second harmonic generation (SHG) for clean semi-conductors have been presented before.\cite{sipe96,sipe93} 
We note that Eq. B3 of Ref \onlinecite{sipe96} is incorrect and the correct form of this equation can be obtained from the sum of Eqs. 
B16b and B17 of Ref \onlinecite{sipe93}. The susceptibility for SHG can be divided into three major contributions: 
the interband transitions $\chi_{\rm inter}(2\omega,\omega,\omega)$, 
the intraband transitions $\chi_{\rm intra}(2\omega,\omega,\omega)$ and 
the modulation of interband terms by intraband terms $\chi_{\rm mod}(2\omega,\omega,\omega)$:

\begin{eqnarray} \label{sipe1}
\chi_{\rm inter}^{abc}(2\omega,\omega,\omega) = 
{{e^3} \over {\hbar^2}} \sum'_{nml} \int {{d\bf k} \over {4 \pi^3}} 
{{r^a_{nm} \{ r^b_{ml}r^c_{ln}\}} \over {(\omega_{ln}-\omega_{ml})}} 
\left\{
{{2f_{nm}} \over {(\omega_{mn}-2\omega)}} +
{{f_{ml}} \over {(\omega_{ml}-\omega)}}+{{f_{ln}} \over {(\omega_{ln}-\omega)}} 
\right\}
\end{eqnarray}

\begin{eqnarray} \label{sipe2}
\chi_{\rm intra}^{abc}(2\omega,\omega,\omega) = {{e^3} \over {\hbar^2}} \int {{d\bf k} \over {4 \pi^3}} 
\left[
\sum'_{nml} \omega_{mn}r^a_{nm}\{r^b_{ml}r^c_{ln}\} 
\left\{ 
{{f_{nl}} \over {\omega^2_{ln}(\omega_{ln}-\omega)}} - {{f_{lm}} \over {\omega^2_{ml}(\omega_{ml}-\omega)}}
\right\} \right. \nonumber \\
\left. -8i\sum'_{nm} {{f_{nm} r^a_{nm}\{\Delta^b_{mn}r^c_{mn}\}} \over {\omega^2_{mn}(\omega_{mn}-2\omega)}} 
+2\sum'_{nml} {{f_{nm} r^a_{nm}\{r^b_{ml}r^c_{ln}\}(\omega_{ml}-\omega_{ln})} \over {\omega^2_{mn}(\omega_{mn}-2\omega)}}
 \right]
\end{eqnarray}

\begin{eqnarray} \label{sipe3}
\chi_{\rm mod}^{abc}(2\omega,\omega,\omega) = {{e^3} \over {2\hbar^2}} \int {{d\bf k} \over {4 \pi^3}} 
\left[
\sum'_{nml} {{f_{nm}} \over {\omega^2_{mn}(\omega_{mn}-\omega)}} 
\left\{
\omega_{nl}r^a_{lm}\{r^b_{mn}r^c_{nl}\}-\omega_{lm}r^a_{nl}\{r^b_{lm}r^c_{mn}\}
\right\}
\right. \nonumber \\
\left. 
-i\sum'_{nm} {{f_{nm}r^a_{nm}\{r^b_{mn}\Delta^c_{mn}\}} \over {\omega^2_{mn}(\omega_{mn}-\omega)}}
\right]
\end{eqnarray}
for all $n \ne m \ne l$. The symbols are defined as
\begin{eqnarray}
\Delta^a_{nm}({\bf k}) = v^a_{nn}({\bf k}) - v^a_{mm}({\bf k})
\end{eqnarray}
with $v^a_{nm}$ being the $a$ component of the electron velocity given as
\begin{eqnarray}
v^a_{nm}({\bf k}) =iw_{nm}({\bf k})r^a_{nm}({\bf k})
\end{eqnarray} 
and
\begin{eqnarray}
\{r^a_{nm}({\bf k}) r^b_{ml}({\bf k})\} = {1 \over 2}(r^a_{nm}({\bf k})r^b_{ml}({\bf k}) + r^b_{nm}({\bf k}) r^a_{ml}({\bf k}))
\end{eqnarray}
The position matrix elements between states $n$ and $m$, $r^a_{nm}({\bf k})$, are calculated from the momentum matrix element $p^a_{nm}$ 
using the relation \cite{cad}
\begin{eqnarray}
r^a_{nm}({\bf k}) = {p^a_{nm}({\bf k}) \over { i m \omega_{nm}({\bf k})}}
\end{eqnarray}
for all $\omega_n({\bf k}) \ne \omega_m({\bf k})$  and $r^a_{nm}({\bf k})$ = 0 otherwise.
with the energy difference between the states $n$ and $m$ given by
\begin{eqnarray}
\hbar \omega_{nm} = \hbar (\omega_n - \omega_m).
\end{eqnarray}
For the sake of clarity the ${\bf k}$ dependence of $\omega_{nm}$, $p^a_{nm}$, $v^a_{nm}$ and $\Delta_{nm}$ in the 
Eqs. (1)-(3) are suppressed. 
Equations (1) to (3) are computationally very demanding since $n, l$ and $m$ run over all the bands. But this requirement can 
be relaxed by adjusting the equations slightly using the fact that the equations are symmetric in $n, l$ and $m$ and hence the indices 
can be interchanged. At the same time one can get rid of the Fermi functions. Eqs. (1)-(3) then read 

\begin{eqnarray}
\chi_{\rm inter}^{abc}(2\omega,\omega,\omega) = 
{{e^3} \over {\hbar^2\Omega}} \sum'_{nml}\sum_{\bf k} W_{\bf k}
\left\{
{{2r^a_{nm}\{r^b_{ml}r^c_{ln}\}} \over{(\omega_{ln}-\omega_{ml})(\omega_{mn}-2\omega)}}
\right. \nonumber \\
\left.
-{1 \over (\omega_{mn}-\omega)} 
\left[
{{r^c_{lm}\{r^a_{mn}r^b_{nl}\}} \over{(\omega_{nl}-\omega_{mn})}}-
{{r^b_{nl}\{r^c_{lm}r^a_{mn}\}} \over{(\omega_{lm}-\omega_{mn})}}
\right]
\right\}
\end{eqnarray}

\begin{eqnarray}
\chi_{\rm intra}^{abc}(2\omega,\omega,\omega) = 
{{e^3} \over {\hbar^2\Omega}} \sum_{{\bf k}} W_{\bf k}
\left\{
\sum'_{nml} {1 \over {\omega^2_{mn}(\omega_{mn}-\omega)}} 
\left[
\omega_{ln}r^b_{nl}\{ r^c_{lm}r^a_{mn}\} -\omega_{ml}r^c_{lm}\{ r^a_{mn}r^b_{nl}\}
\right]
\right. \nonumber \\
\left.
-8i\sum'_{nm} {1 \over{\omega^2_{mn}(\omega_{mn}-2\omega)}}r^a_{nm}\{ r^b_{ml}r^c_{ln}\}
+2\sum'_{nml}{{r^a_{nm}\{ r^b_{ml}r^c_{ln}\}(\omega_{ml}-\omega_{ln})} \over {\omega^2_{mn}(\omega_{mn}-2\omega)}}
\right\}
\end{eqnarray}

\begin{eqnarray}
\chi_{\rm mod}^{abc}(2\omega,\omega,\omega) = {{e^3} \over {2\hbar^2\Omega}} \sum_{{\bf k}} W_{\bf k}
\left\{
\sum_{nml} {1 \over {\omega^2_{mn}(\omega_{mn}-\omega)}} 
\left[
\omega_{nl}r^a_{lm}\{r^b_{mn}r^c_{nl}\}-\omega_{lm}r^a_{nl}\{r^b_{lm}r^c_{mn}\}
\right]
\right. \nonumber \\
\left. 
-i\sum_{nm} {{r^a_{nm}\{r^b_{mn}\Delta^c_{mn}\}} \over {\omega^2_{mn}(\omega_{mn}-\omega)}}
\right\}
\end{eqnarray}
where $\Omega$ is the unit cell volume, $W_{\bf k}$ is the weight of $k$ point and $n$ denotes the valence states, $m$ the conduction states 
and $l$ denotes all states ($l \ne m, n$). We have used the Eqs. (9)-(11) to calculate the total susceptibility. We also confirm 
that the real and imaginary parts of this susceptibility satisfy all the sum rules presented by Scandolo {\it et al.} \cite{scandolo95} 
Some results of the benchmark are presented in the following table. All the calculations are performed with
a mesh of 1000 points in the energy interval 0.0 - 4.0 eV on a SGI R12000 processor.

\begin{table}[h]
\caption{Number of $k$-points in the IBZ, number of valence bands (VB), number of conduction bands (CB), CPU time (in minutes) needed to calculate the 
SHG susceptibility using Eqs. (1)-(3) and Eqs. (9)-(11).}
\centerline{
\begin{tabular}{|c| c  c | c c| } \hline
$k$ points &  Number of & bands    & CPU           & time         \\ 
IBZ        &   VB      &  CB       & Eqs. (1)-(3)  & Eqs. (9)-(11) \\ \hline
385        &  9        &   11      &   11.02       &   2.67        \\ \hline
385        &  9        &   5       &   4.50        &   1.10        \\ \hline
146        &  9        &   11      &   3.50        &   0.88        \\ \hline
146        &  24       &   33      &   88.01       &   19.47       \\ \hline
\end{tabular}
}
\end{table}

\section{Appendix B: FORMALISM FOR THE STRAIN-CORRECTED EFFECTIVE MEDIUM MODEL}

The expression for calculating the $xx$ component of the linear optical response using the effective medium model is:\cite{ghahramani90}
\begin{eqnarray}
\epsilon^{eff}_T(\omega,\delta,\Delta) = {1 \over 2}  (\epsilon_A^{xx}(\omega,\delta) + \epsilon_B^{xx}(\omega,\Delta))
\end{eqnarray}
which can be obtained from the continuity of the tangential electric field across the interface $E_T = E^A_T$ and $ E_T = E^B_T$ and the 
expression for the average induced polarization $P^{eff}= {1 \over 2} (P^A+P^B)$. The longitudinal electric field on the other hand is given by
$\epsilon^{zz}_{eff} E_L = \epsilon^{zz}_A E_L^A = \epsilon^{zz}_B E_L^B $, and so the expression for $\epsilon_{zz}^{eff}$ is given by
\begin{eqnarray}
\epsilon^{eff}_{zz}(\omega,\delta,\Delta) = {{2 \epsilon_A^{xx}(\omega,\delta) \epsilon_B^{xx}(\omega,\Delta)} \over 
{ \epsilon_A^{xx}(\omega,\delta) + \epsilon_B^{xx}(\omega,\Delta)}}
\end{eqnarray}
Here it is  assumed that A is the substrate material and B is the material grown on top of A and hence has the strained lattice parameters.
$\epsilon_A^{xx}(\omega,\delta)$ is the $xx$ component of the frequency dependent dielectric tensor for material A (or strained material B) 
corrected with the scissors operator $\delta$ (or $\Delta$).  Where $\delta$ and $\Delta$ are calculated as:

\begin{eqnarray}
\delta = XDG - TOG_{X}  
\end{eqnarray}

\begin{eqnarray}
\Delta = XDG  - TOG_{S}  
\end{eqnarray}
where XDG is the experimental direct band gap and TOG is the optical bang gap calculated theoretically. The subscripts X and S indicate that the 
experimental lattice parameters and the strained lattice parameters, respectively, are used for calculating the theoretical optical band gap. The EMM 
assumes that even after straining the material B has the same gap as measured experimentally for the unstrained material. 
These expression are slightly altered for the SCEMM reading:

\begin{eqnarray}
\epsilon^{eff}_T(\omega,\delta) = {1 \over 2}  (\epsilon_A^{xx}(\omega,\delta_A) + \epsilon_B^{xx}(\omega,\delta_B))
\end{eqnarray}

\begin{eqnarray}
\epsilon^{eff}_{zz}(\omega,\delta) = {{2 \epsilon_A^{xx}(\omega,\delta_A) \epsilon_B^{xx}(\omega,\delta_B)} \over 
{ \epsilon_A^{xx}(\omega,\delta_A) + \epsilon_B^{xx}(\omega,\delta_B)}}
\end{eqnarray}
where the symbols have the usual meaning. One extra set of calculations is needed for the unstrained B material to 
determine the optical band gap and hence the scissors operator. Then the calculations are performed for the strained
material B and the band gap is corrected with the ``old'' scissors operator. This effective $zz$ component of the dielectric tensor can be used to 
calculate the  $xyz$ component of the SHG susceptibility as follows:

\begin{eqnarray}
\chi_{xyz}^{eff}(2\omega,\omega,\omega,\delta) = 
{{\epsilon^{eff}_{zz}(\omega, \delta)} \over 2} \left[ 
{{\chi_{xyz}^A(2\omega,\omega,\omega,\delta_A)} \over {\epsilon_{zz}^A(\omega,\delta_A)}} +
{{\chi_{xyz}^B(2\omega,\omega,\omega,\delta_B)} \over {\epsilon_{zz}^B(\omega,\delta_B)}}
 \right].
\end{eqnarray}

\noindent \textbf{Acknowledgements}

We would like to thank the Austrian Science Fund (project P13430) for  financial support.
The code for calculating non-linear optical properties was written under the EXCITING network funded by the EU (Contract HPRN-CT-2002-00317).
SS would also like to thank Bernd Olejnik for useful discussions and suggestions during the course of code development.  


\end{document}